\documentclass[twocolumn,showpacs,preprintnumbers,amsmath,amssymb]{revtex4}
\usepackage{graphicx}
\usepackage{dcolumn}
\usepackage{bm}
\pdfoutput=1

\begin{document}
\title{Topological entanglement entropy of interacting disordered zigzag graphene ribbons}

\author{Young Heon Kim}
\thanks{These two authors contributed equally}
\author{Hye Jeong Lee}
\thanks{These two authors contributed equally}
\author{S.-R. Eric Yang}
\email[corresponding author:\;]{eyang812@gmail.com}
\affiliation{Department of Physics, Korea University, Seoul, Korea}

\begin{abstract}
Interacting disordered zigzag graphene nanoribbons have fractional charges, are quasi-one-dimensional, and display an exponentially small gap. Our numerical computations showed that the topological entanglement entropy of these systems has a small finite but universal value, independent of the strength of the interaction and  the disorder. The result that was obtained for the topological entanglement entropy shows that the disorder-free phase is critical and becomes unstable in the presence of disorder.
 Our result for the entanglement spectrum in the presence of disorder is also consistent with the presence of a topologically ordered phase.
\end{abstract}
\maketitle

\section{Introduction}

\begin {figure}[!hbpt]
\begin{center}
\includegraphics[width=0.3\textwidth]{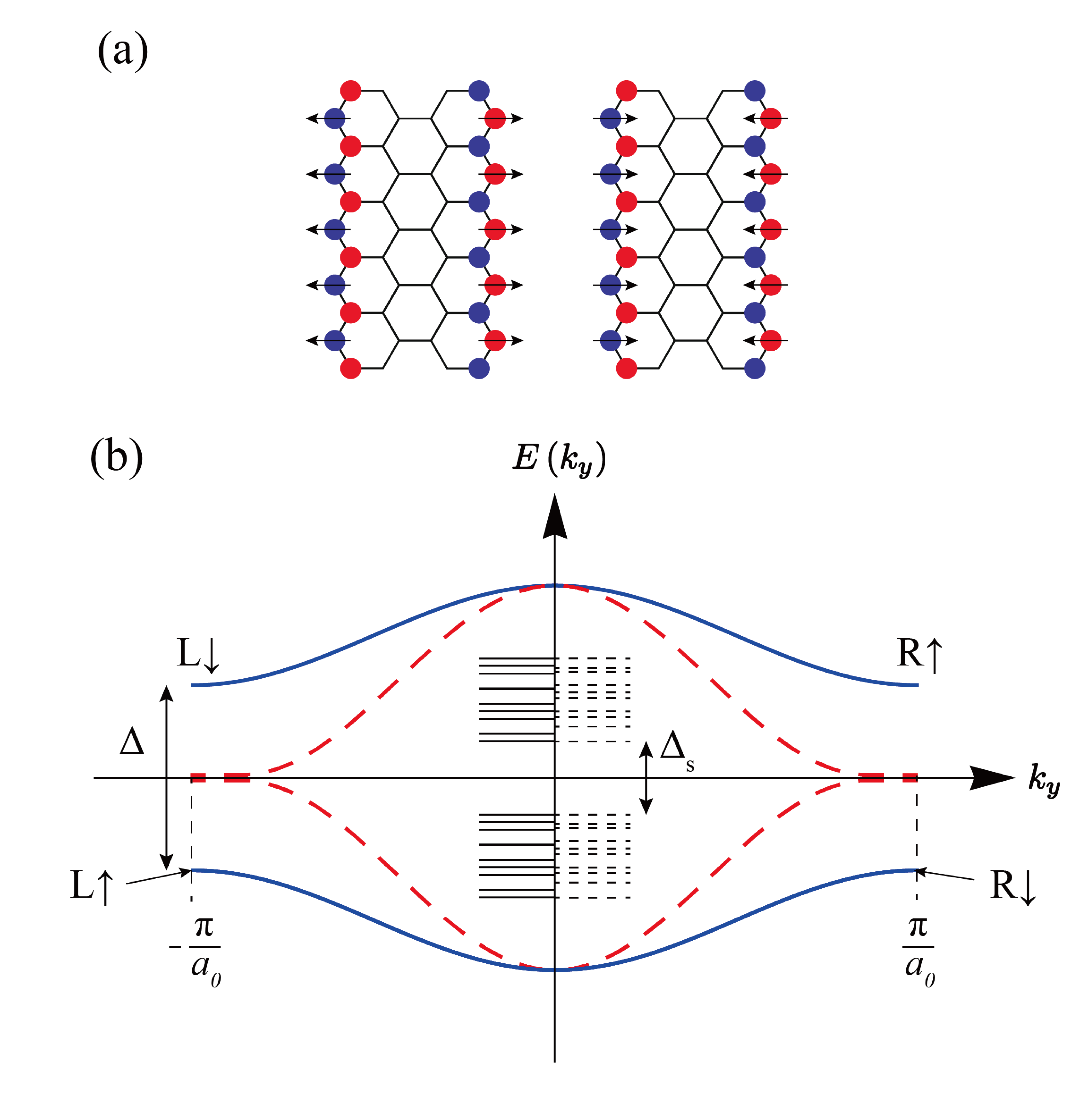}
\caption{(a) Zigzag edge antiferromagnetism of an interacting zigzag graphene nanoribbon without disorder showing the two degenerate ground states. (b) Schematic band structures of interacting (solid curves) and non-interacting (dashed curves) zigzag graphene nanoribbons. Unoccupied and occupied  
states near the wave vectors $k=\pm \pi/a_0$: $R$ and $L$ represent the states confined to the zigzag edges on the right and left, respectively (the length of the unit cell of a ribbon is $a_0$).  
The small arrows indicate the spins. Spin-split energy levels of the spin-up (solid lines) and spin-down (dashed lines) {\it gap--edge} states of the interacting disordered interacting zigzag graphene nanoribbons. These figures are taken from Ref. \cite{Yang}.
}
\label{degGap}
\end{center}
\end{figure}

\begin {figure}[!hbpt]
\begin{center}
\includegraphics[width=0.25\textwidth]{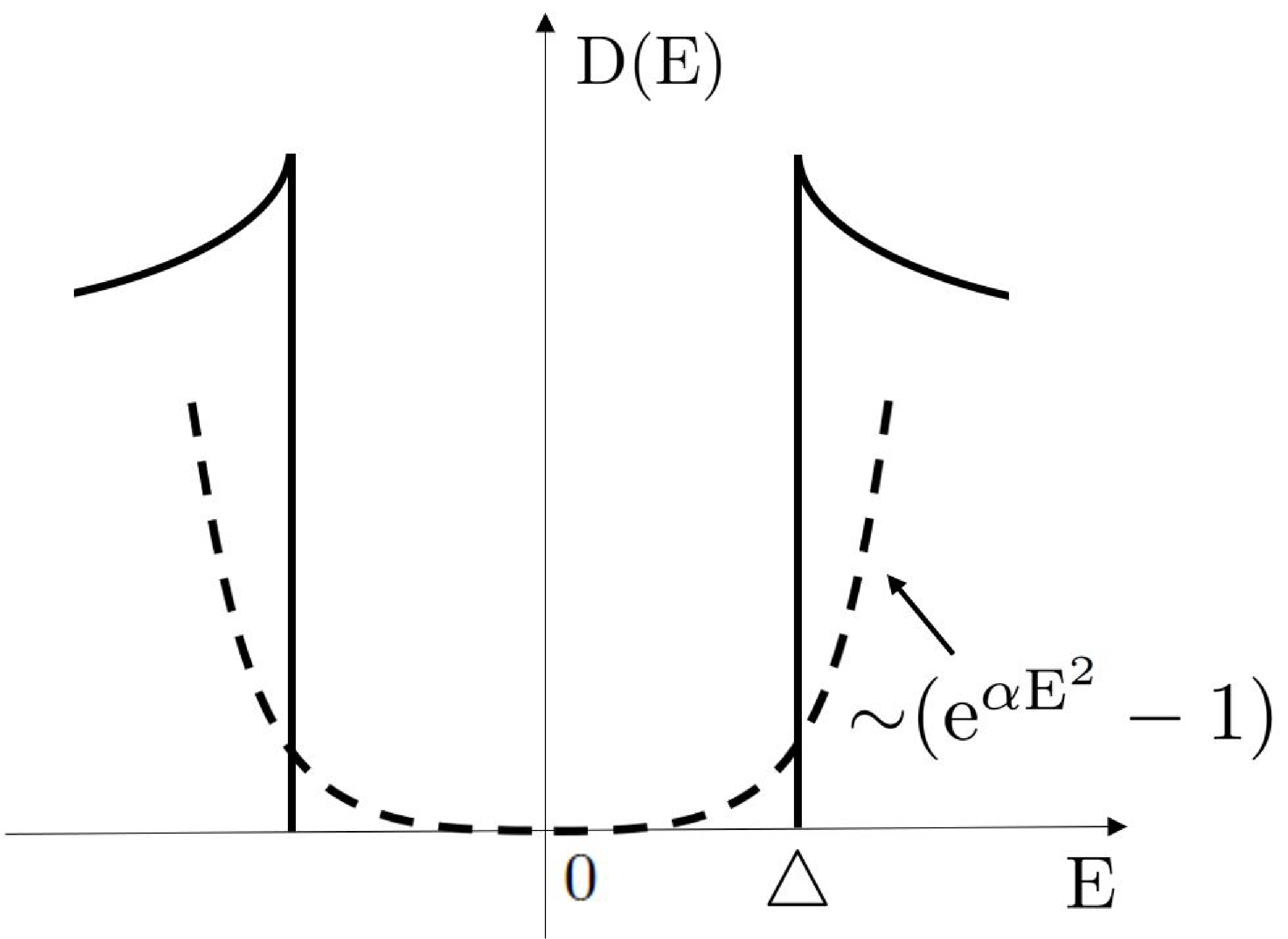}
\caption{Dashed line: the exponentially small {\it soft} gap with  $\alpha\sim 1/\sqrt{\Delta_s}$ near the Fermi energy  of the tunneling density of states of an interacting disordered zigzag graphene nanoribbon.  Solid line:  the {\it hard} gap, $\Delta$, of the tunneling density of states of an interacting zigzag graphene nanoribbon in the absence of disorder.}
 \label{Half}
\end{center}
\end{figure}

Graphene structures have many fascinating physical properties, such as massless Dirac electrons and the quantum Hall effect \cite{Nov,Zhang,Neto,Geim}. Recent years have seen rapid progress in the fabrication of atomically precise graphene nanoribbons \cite{Cai2}.
Recent studies showed \cite{Jeong,Yang} that interacting  zigzag graphene nanoribbons~\cite{Fujita} are topologically ordered ~\cite{Wen1,Wen2} when disorder is present. The ground states of these quasi-one-dimensional Mott-Anderson insulators are doubly degenerate: the two degenerate ground states are related to each other in that their electron spins are reversed (see Fig. \ref{degGap}). They have an exponentially small gap, $\Delta_s$, in the DOS (see Fig. \ref{Half}) and their  boundary zigzag edges can support $1/2$ fractional charges. These objects are solitonic in nature \cite{Jeong1,Brey}. Moreover, the zigzag edges induce spin-splitting in the bulk~\cite{Jeong2} and can display spin-charge separation.
Disorder in interacting zigzag graphene ribbons induces a transition between symmetry-protected and topologically ordered phases.

In this study, we investigated the entanglement in many-body topological insulators \cite{Ami,GY} of interacting disordered zigzag graphene nanoribbons.  A region $D$ of a topologically ordered {\it gapful} system with {\it one} boundary has an entanglement 
entropy \cite{Eis,Wen,Pac} 
\begin{eqnarray}
S_D=\alpha L-\beta
\label{TopoEntropy}
\end{eqnarray}
with the  sub-dominant and universal  topological entanglement entropy (TEE) $\beta>0$ \cite{Kitaev,Levin}.
The value of  $\beta$ can be numerically computed, see, for example, Jiang et al. \cite{Bal}.  Here $L$ is the length of the boundary and $\alpha$ is non-universal constant. The entanglement entropy $S_D=-\text{Tr}[\rho_D \text{ln}\rho_D]$  is given by the reduced density matrix $\rho_D$ of region $D$.

\begin {figure}[!hbpt]
\begin{center}
\includegraphics[width=0.45\textwidth]{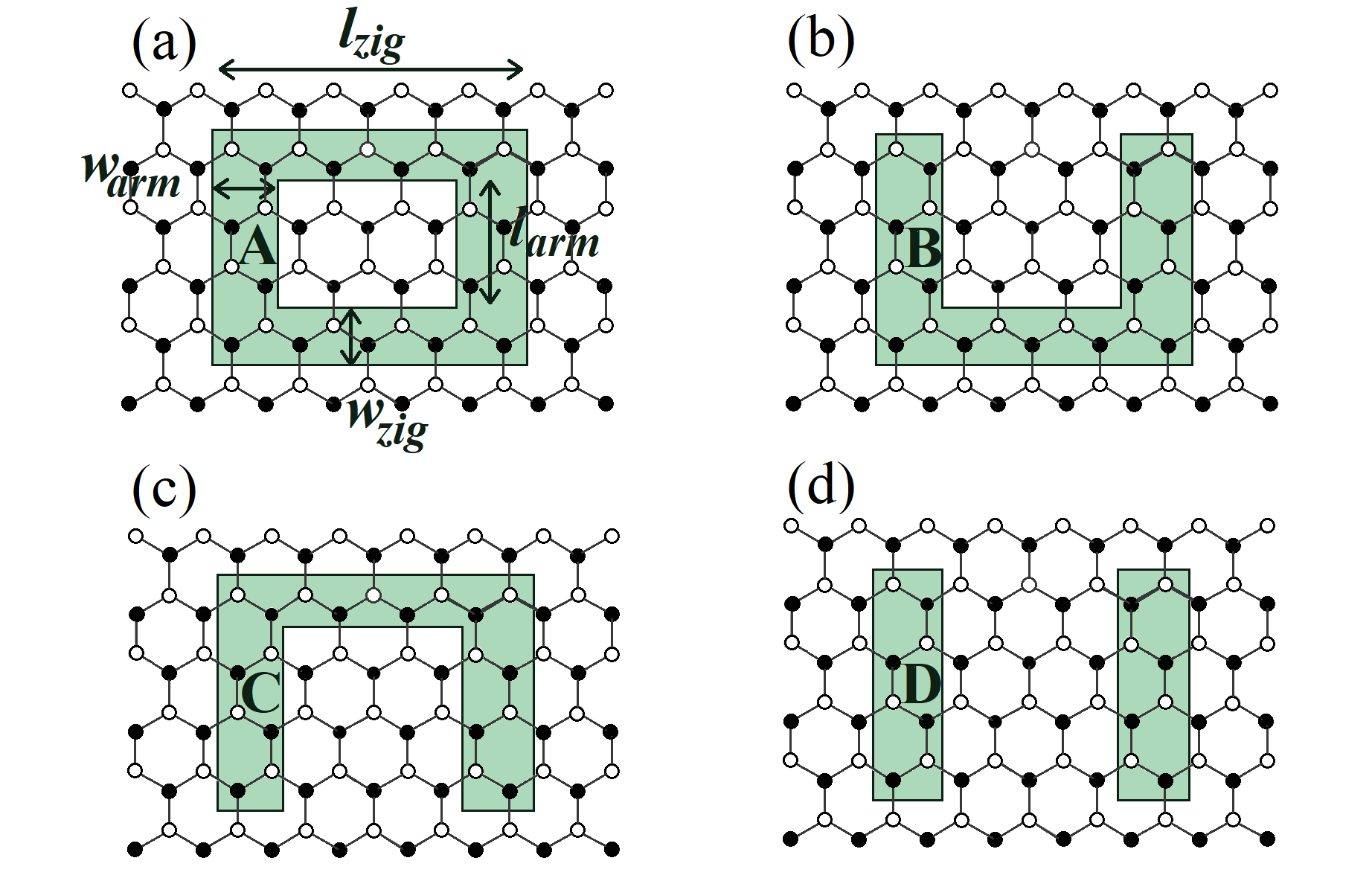}
\caption{ Different regions of the ribbon used in Eq.(\ref{Entan})  to compute the TEE.  Vertical lengths  are measured in   number of  horizontal carbon lines and  horizontal lengths in   number of   vertical  carbon lines. }
\label{Ring1}
\end{center}
\end{figure}

In the case of non-negligible finite-size effects and disorder fluctuations, different methods can be used to compute $\beta$. 
We adopt the  method of partitioning the system into different regions \cite{Kitaev,Levin}.  The TEE  of a ring\cite{Levin} can be written as
\begin{eqnarray}
S_{top}=2\beta=-[(S_{A}- S_{B}) -(S_{C}-S_{D})].
\label{Entan}
\end{eqnarray}
 $S_{A}$ is the  entanglement  entropy for the region consisting of the sites in
$A$.  Other entanglement  entropies  defined similarly.   The regions should be as in Fig. \ref{Ring1}  or a smooth deformation
thereof without changing how the regions border on each other.  
The entanglement entropy $S_{A}-S_{B}$ has the contribution as that of  $S_{C}-S_{D}$. The difference between these two contributions  is the TEE.   Note that other regions have only one boundary, whereas
$A$ has {\it two} boundaries, i.e., one inner and one outer, as shown in Fig. \ref{Ring1}. Except near a {\it critical point}, the value of  the TEE   is universal and is independent of system parameters.

Several questions regarding the entropy of interacting disordered zigzag ribbons remain unanswered. For example, it is unclear whether these ribbons should have non-zero TEE. As mentioned above, the presence of fractional charges suggests that the TEE   is finite, although the magnitude thereof is unknown. However, the presence of an exponentially small gap may be compatible with zero TEE. In addition, the behavior of the TEE  near the critical point is unclear.
In this work, we numerically computed the TEE  and showed that it is finite and universal, and independent of the strength of both the interaction and disorder. Our results also showed that the disorder-free phase is critical, and that this phase becomes unstable in the presence of disorder.

\section{Model}

We applied a Hubbard model to the interacting disordered zigzag graphene nanoribbons and 
used a self-consistent Hartree-Fock  approximation \cite{GY,Mac1}. 
We included both electron–electron interactions and disorder in a tight-binding model at half-filling.
When the on-site repulsion is $U=0$, the effect of disorder can be described exactly  within the Hartree-Fock  approximation, whereas in the other limit, where disorder is absent, the interaction effects are well represented by the Hartree-Fock  approximation, which is widely used in graphene-related systems \cite{Stau}.  
The ground state is doubly degenerate and can be written as a product of  spin-up and -down Slater determinants:
\begin{eqnarray}
\Psi_1&=&\Psi_{L,\uparrow}(\vec{r}_1,\ldots,\vec{r}_{N/2})\Psi_{R,\downarrow}(\vec{r}_{N/2+1},\ldots,\vec{r}_{N}),\nonumber\\
\Psi_2&=&\Psi_{R,\uparrow}(\vec{r}_1,\ldots,\vec{r}_{N/2})\Psi_{L,\downarrow}(\vec{r}_{N/2+1},\ldots,\vec{r}_{N}),
\end{eqnarray}
where $\Psi_{L,\sigma}$ ($\Psi_{R,\sigma}$) describes 
 $N/2$ electrons with  spin $\sigma$ with some of these  electrons  localized on the left (right) zigzag edge.  
In the first state the spin of the magnetization of the zigzag edge on the left (right) is dominantly upward (downward), see Fig. \ref{degGap}(a). In the second state, the magnetization is the opposite.  The total number of electrons is $N$.    In these wave functions spins are separated because spins terms of  the mean field Hamiltonian are separated,   $H=H_{\uparrow}+H_{\downarrow}$ \cite{Jeong,Yang},   see the following equation.

The total Hamiltonian in the Hartree-Fock approximation  is
\begin{eqnarray}
&&H=-\sum_{<ij>\sigma} t c_{i\sigma}^{\dag}c_{j\sigma}+\sum_{i\sigma} V_ic_{i\sigma}^{\dag}c_{i\sigma}\nonumber\\
&+&U\sum_{i}  (n_{i\uparrow} \langle n_{i\downarrow}\rangle+\langle n_{i\uparrow}\rangle n_{i\downarrow}
-\langle n_{i\uparrow}\rangle \langle n_{i\downarrow}\rangle )\nonumber\\
&-&\frac{U}{2}\sum_i (n_{i\uparrow} +n_{i\downarrow} ),
\label{Ham}
\end{eqnarray}
where $c_{i\sigma}^{\dag}$ and $n_{i\sigma}$ are the electron creation and  occupation operators at site $i$ with spin $\sigma$. Because the translational symmetry is broken, the Hamiltonian is written in the site representation. In the hopping term, the summation is over the nearest neighbor sites (the  value of the hopping parameter is $t\sim 3 eV$).
The eigenstates and eigenenergies are computed numerically by solving the tight-binding Hamiltonian matrix self-consistently.  
The self-consistent occupation numbers $\langle n_{i\sigma}\rangle$ in the Hamiltonian are
the sum of the probabilities of finding electrons of spin $\sigma$ at site $i$:
\begin{eqnarray}
\langle n_{i\sigma}\rangle=\sum_{E\leq E_F} |\psi_{i\sigma} (E) |^2 .
\label{occnum}
\end{eqnarray}
The summation is over the energy $E$ of the occupied eigenstates below the Fermi energy $E_F$. 
Note that $ \psi_{i\sigma}(E) $ 
represents an eigenvector of the tight-binding Hamiltonian matrix with  energy $E$.  The on-site impurity energy $V_i$ is chosen randomly from the energy interval $[-\Gamma,\Gamma]$.

\section{Calculation of topological entanglement entropy}

The upward or downward spin destruction operator of a Hartree-Fock single-particle state $|k\rangle$ is
\begin{eqnarray}
a_k=\sum_i {\bf A}_{ki}c_i,
\end{eqnarray}
where $c_i$ is the destruction of the operator at site $i$. Inverting this relation, we find
\begin{eqnarray}
c_i=\sum_k ( {\bf A}^{-1})_{ik}a_k=\sum_k  {\bf B}_{ik}a_k.
\end{eqnarray}

Let us divide the zigzag ribbon into two parts $A$ and $B$. 
We restrict the indices  $i$ and $j$ to $ A$  and define the correlation function/reduced density matrix \cite{Peschel115,Lat}
\begin{eqnarray}
{\bf C}_{ij}&=&\langle \Psi|c_i^+c_j|\Psi\rangle,
\end{eqnarray}
where $ \Psi= \Psi_1$ is chosen ($\Psi_2$ is an equally good choice). This can be written as
\begin{eqnarray}
{\bf C}_{ij}&=&\langle \Psi|(\sum_k  {\bf B}_{ik}a_{k})^{+}(\sum_{k'} {\bf B}_{jk'}a_{k'})|\Psi\rangle\nonumber\\
&=&\langle \Psi|(\sum_k a_{k}^+ {\bf B}^{*}_{ki})(\sum_{k'} {\bf B}_{jk'}a_{k'})|\Psi\rangle\nonumber\\
&=&\sum_k  {\bf B}^{*}_{ki} {\bf B}_{jk}n_k.
\end{eqnarray}
Here we have used 
\begin{eqnarray}
\langle \Psi|a_k^+a_{k'}|\Psi\rangle&=&\delta_{kk'}n_k,
\end{eqnarray}
where $n_k$ is the number of occupied Hartree-Fock states.  
The entanglement entropy of $A$ is given by
\begin{eqnarray}
S_A=-\sum_{i}\left[ \lambda_{i}\text{ln}\lambda_{i}
+(1-\lambda_{i})\text{ln}(1-\lambda_{i})\right],
\end{eqnarray}
where $\lambda_i$ are the eigenvalues of the matrix $\bf{C}$.

\section{ Topological entanglement entropy }

We computed $S/L$ as a function of $1/L$ and tried to extract the TEE   from the result. 
 For an interacting but disorder-free system, we found $\beta\approx 0$. Note that, in this case, a hard gap exists in the DOS and {\it no} zero-energy zigzag edge states exist, as shown in Fig. \ref{Half}.
However, 
in the presence of disorder, the method for computing the TEE  from Eq.(\ref{TopoEntropy}) is not accurate: the data we obtained display significant disorder fluctuations that depend on the size of the system.

We use another method based on a rectangular ring   to compute $S_{top}=2\beta$ given in Eq.(\ref{Entan}).   Since the whole  ribbon is topologically ordered \cite{Jeong,Yang} we can  divide the system into  a rectangular ring that lies well inside the ribbon and the rest of the ribbon.  
However, 
a careful analysis is required to reduce  finite-size effects, see Jiang et al. \cite{Bal}.   For this purpose we adopt the  partition of the ribbon  shown in  Fig.\ref{Ring1} \cite{Levin}.  
We apply this method to compute $\beta$ of interacting disordered zigzag ribbons.  To compute  $\beta$  one must use a  thick and large rectangular ring. 
The width of the ring $w=w_{zig}=w_{arm}$ should be   larger than the correlation length $\xi$. (In zigzag ribbons the correlation length $\xi\sim\hbar v_F/\Delta$ is of the order of the lattice constant, estimated from the size of gap $\sim 2$eV ~\cite{Fujita}  and the Fermi velocity $v_F$).    When the width increases  the size of  ribbon and the ring  must be increased at the same time.
As a  check on our method we  computed the  TEE  of a gapful armchair nanoribbon.  We  found the expected value of zero both in the absence and presence of disorder.

\begin{figure}[!hbpt]
\begin{center}
\includegraphics[width=0.45\textwidth]{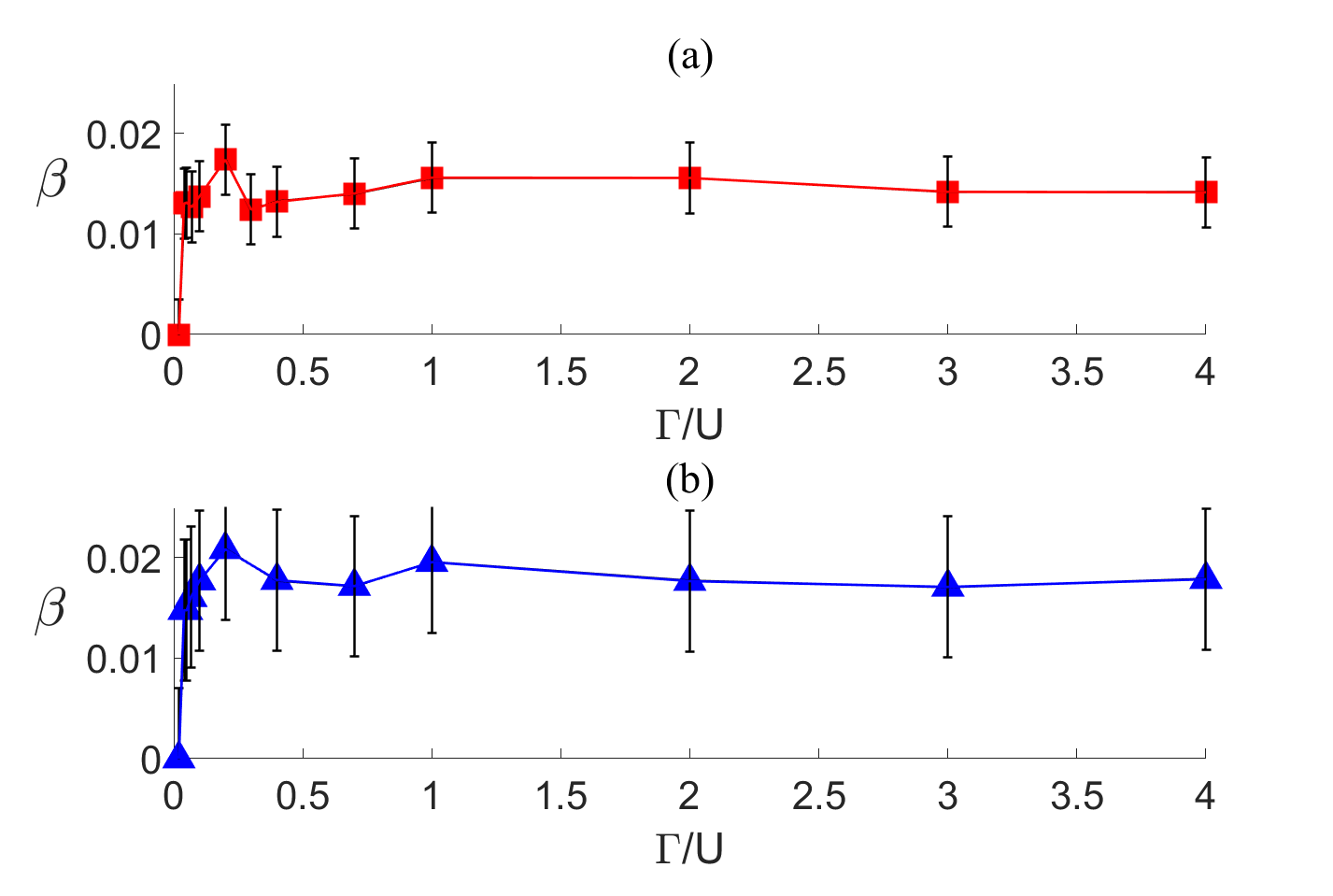}
\caption{TEE  $\beta$ of an interacting disordered  zigzag nanoribbon
is computed using  a rectangular ring with different ring widths $w$.     Each point represents a different  value of $(U,\Gamma)$.
In (a)  the ring width, ribbon length, ribbon width are, respectively, $w=7$, $L'=260$, $W=112$. The side lengths of the ring are $l_{zig}=221$ and  $l_{arm}=46$.  In  (b) they are $w=4$, $L'=200$      $W=64$, $l_{zig}=161$, $l_{arm}=28$.  The number of disorder realizations $N_D$ are, respectively, $10$ and  $20$ in (a) and (b).}
\label{DataCollp}
\end{center}
\end{figure}

The TEE result  of  interacting disordered  zigzag graphene nanoribbons   is
displayed in Fig. \ref{DataCollp} for two values of the ring width.   We see that, as the ring width increases, the numerical 
uncertainties decrease.  We find   that the numerical uncertainty of the computed TEE is small  when    the ring width is larger than $\gtrsim 7$ (it is estimated by varying the ring width at the fixed values of  the ribbon length and width  $L'\sim 300$ and  $W\sim 100$ and   the rectangular  ring  sizes  $l_{zig}\sim 250$ and    $l_{arm}\sim 50$).  

\noindent
Except for small values of $\Gamma/U$, the results are independent of different values of 
$L'$, $W$, $N_{imp}$, $l_{arm}$, and $l_{zig}$.
From the data collapse  we infer that     $\beta\approx 0.016\pm 0.003$ and that it  is,  within numerical uncertainty, {\it independent} of the strength of the interaction and  disorder, that is,  independent of $\Gamma/U$.  We consider this small value of $\beta$ to be related to the presence of an exponentially small soft  gap in interacting disordered ribbons \cite{Yang}.    We see from the data that, as the critical point is approached, $\Gamma/U\rightarrow 0$, the value of the TEE  varies rather  abruptly   to zero in a non-universal manner \cite{Wen}. (In limit of infinitely large systems   the transition  should occur discontinuously  at $\Gamma/U=0$.)

In conclusion, our numerical work showed that the TEE of interacting disordered zigzag graphene nanoribbons is small but finite and universal. Disorder-free interacting zigzag graphene nanoribbons are in a critical phase that becomes unstable in the presence of disorder.     It would be interesting to find other systems with a pseudo gap that belongs to the same  universality class.    It may be worthwhile to analytically compute $S_{top}$ of a quasi-one-dimensional system in the presence of an exponentially small gap.   

\section*{Acknowledgments}
This research was supported by the Basic Science Research Program 
through the National Research Foundation of Korea (NRF), funded by the
Ministry of Education, ICT $\&$ Future Planning (MSIP) (NRF-2018R1D1A1A09082332).

\end{document}